\documentclass[a4paper,fleqn,usenatbib]{mnras}

\usepackage{newtxtext,newtxmath}

\usepackage[T1]{fontenc}
\usepackage{ae,aecompl}

\usepackage{graphicx}	
\usepackage{amsmath}	
\usepackage{amssymb}	
\usepackage{pdflscape}	

\usepackage{multirow}

\title[Proper motion separation of Be stars]{Proper motion separation of Be star candidates in the Magellanic Clouds and the Milky Way}

\author[K. Vieira]{Katherine Vieira,$^{1}$\thanks{E-mail: kvieira@cida.gob.ve}
Alejandro Garc\'ia-Varela$^{2}$
and Beatriz Sabogal$^{2}$
\\
$^{1}$Centro de Investigaciones de Astronom\'ia, Apartado Postal 264, Merida 5101-A, Venezuela\\
$^{2}$Universidad de los Andes, Departamento de F\'{\i}sica, Cra. 1 \#18A-10, Bloque Ip, AA 4976, Bogot\'a, Colombia
}

\date{Accepted 2017 April 26. Received 2017 April 26; in original form 2017 March 20}

\pubyear{2017}

\begin{document}
\label{firstpage}
\pagerange{\pageref{firstpage}--\pageref{lastpage}}
\maketitle

\begin{abstract}

We present a proper motion investigation of a sample of Be star candidates towards the Magellanic Clouds, 
which has resulted in the identification of separate populations, in the Galactic foreground and in the Magellanic background.  
Be stars are broadly speaking B-type stars that have shown emission lines in their spectra. 
In this work, we studied a sample of 2446 and 1019  Be star candidates towards the LMC and SMC respectively, 
taken from the literature and proposed as possible Be stars due to their variability behaviour in the OGLE-II I band.
JHKs magnitudes from the IRSF catalog and proper motions from the SPM4 catalog, 
were obtained for 1188 and 619 LMC and SMC Be stars candidates, respectively.
Color-color and vector-point diagrams were used to identify different populations among the Be star candidates.
In the LMC sample, two populations with distinctive infrared colours and kinematics were found, 
the bluer sample is consistent with being in the LMC and the redder one with belonging to the Milky Way disk.
This settles the nature of the redder sample which had been described in previous publications as a possible unknown subclass
of stars among the Be candidates in the LMC. In the SMC sample, a similar but less evident result was obtained, since this
apparent unknown subclass was not seen in this galaxy. We confirm that in the selection of Be stars by their variability,
although generally successful, there is a higher risk of contamination by Milky Way objects towards redder B$-$V and V$-$I colors.
\end{abstract}




\section{Introduction}

Be stars are broadly defined as non-supergiant (luminosity class III to V) B-type stars that have 
or have had Balmer emission lines \citep{collins}.
The presence of a flattened circumstellar gaseous disk formed of material ejected from the star,
a dust-free Keplerian {\it decretion disc}, is currently the accepted explanation for some of the observed features in Be stars:
the UV stellar light is reprocessed in it and produces the emission lines, and the observed
IR excess and polarization result from the scattering of the stellar light by the disk (see \citet{review13} for details).
Several mechanisms have been proposed for the mass-ejection process that forms the disk, 
which are well constrained but not totally understood. In the so-called Classical Be 
it clearly comes from the rapid rotation of the star, 
probably along with other processes including non-radial pulsations and small-scale magnetic fields;
and in binaries stars the material is being accreted
by the companion of the Be star, generally a white dwarf. 

\citet{mennickent02} and \citet{sabogal05}, have identified a large number of Be star candidates 
towards the Magellanic Clouds based on their photometric variability in the OGLE-II I band.
\citet{mennickent02} has proposed a classification of these variability-selected stars into four types
according to their light curves morphology and suggest that 
Type-1 and Type-2 stars could be Be stars with
accreting white dwarfs in a Be + WD binary or blue
pre-main-sequence stars showing accretion disc thermal instabilities,
Type-4 stars could be single Be stars and Type-3 stars should not
be linked to the Be star phenomenon at all. 
They found also a sample of transition stars between Type-1 and Type-2 stars, 
and suggested that these stars could be experiencing  the same phenomenon, 
which could be outbursts of plasma,  accretion from a  white dwarf companion, 
or accretion disc thermal instabilities like these found in blue pre-main sequence stars. 
This last hypothesis was discarded in the study by \citet{mennickent09}.

\citet{paul12} showed that the photometric method used in the aforementioned works
is effective in the selection of Be star candidates. Their spectroscopic analysis found that most of the stars studied from a sample of such candidates
in both LMC and SMC, belong to early type stars with emission supporting circumstellar material.
However an enigmatic result was obtained in their work among Type-4 LMC Be candidate stars:
a subgroup of the brightest and most massive stars was found in the sample
with no NIR excess and large reddening although they were not located
in regions with high reddening. Stars with similar characteristics were not found in the SMC 
or in our Galaxy and it was proposed they formed a possible subclass of stars that needed further analysis.

In this work we used the Southern Proper Motion 4 catalogue (SPM4) \citep{spm4} to study the kinematics of
Be star candidates of the LMC and SMC galaxies, to probe the presence of contaminants or various populations in the studied samples,
and with such information gain a better insight into the nature of these stars.
This also allowed us to independently evaluate how successful the photometric variability techniques used by \citet{mennickent02}
and \citet{sabogal05} are to select Be star candidates, and confirm that some biases could be present in such method.

\section{Cross-matching Be stars with IRSF and SPM4} 

Be star candidates for the LMC and SMC were obtained from \citet{sabogal05} and  \citet{mennickent02},
respectively. A total of 2446 and 1019 candidates were listed, but  we found
four  and two  repeated entries in each catalog. In the LMC, the four sources that appeared twice each,
had the same sky coordinates, and very similar magnitudes and colors, only differing in their Type.
In each case, one entry was a Type-4, and for the lack of better information, we kept that entry in the
catalog and deleted the other. In the SMC, a similar situation was faced and since in the two cases one entry was a Type-1 star,
this one was selected in each case. The final lists of unique entries of Be star candidates  to match contained 2442 and 1017 stars in the LMC
and SMC, respectively.

A first match by sky coordinates was done between the Be star candidates and the SPM4 catalog, with 3"
as a matching radius. A few multiple matches (two Be stars matched to the same SPM4 star and viceversa),
were found. Since SPM4 measured a V magnitude, a comparison with their OGLE-II's V
could provide additional information to clarify these problematic matches and also to discard
false matches, but the SPM4 photometry in the LMC and SMC areas comes mostly from photographic plates
and has a poor quality compared to CCD photometry. We also noticed the J-H vs. H-K color-color diagram
done with the JHKs 2MASS magnitudes listed by SPM4, suffered from a noticeable higher dispersion,
as compared to those published by \citet{paul12} using the InfraRed Survey Facility (IRSF) magnitudes by \citet{kato07}.

Therefore, we decided to make a cross-match between SPM4 and IRSF in the areas that cover the location
of the Be star candidates in the LMC and SMC, respectively, to obtain a SPM4$\times$IRSF catalog with proper motions
and good infrared magnitudes. Then a second match between the Be star candidates and this SPM4$\times$IRSF catalog was performed
to make the final cross-identifications. While the IRSF photometry is of better quality, the catalog contains 
a non-negligible portion of false or poor quality entries (extended/faint/saturated/etc. sources), therefore in an effort
to consider only good quality point sources, we selected only those with the quality and periphery flags both
equal to $111$ and the adjacency flag equals to $000$. By applying such restrictions, the IRSF catalog was reduced
to about 15\% of the original total number of listed entries. To our surprise, we found a few tens of these point sources
with repeated IDs, that had the same sky coordinates and similar photometry (within 0.1-0.2 magnitudes),
and for the lack of better information, we chose the first one to appear for each case.

As IRSF has a fainter limiting magnitude than SPM4, the risk for mismatches between the catalogs by using
only the position in the sky cannot be ignored. But SPM4 2MASS-measured JHKs magnitudes help us to overcome this 
trouble, by looking also for a similarity in the J-band photometry. 
We first transformed the 2MASS J magnitude listed in SPM4 to the IRSF system by applying the corresponding
transformation, listed in Table 10 of \citet{kato07} as 
J\textsubscript{IRSF}=J\textsubscript{2MASS}-0.043(J\textsubscript{2MASS}-H\textsubscript{2MASS})+0.018.
Then, for all IRSF sources within 3" in the sky and within 0.5 magnitudes in J magnitude from a SPM4 one,
a match is done with the IRSF star with the smallest
\begin{equation}
d=\sqrt{\left(\frac{\Delta\theta}{3"}\right)^2+\left(\frac{\Delta\mbox{J}}{0.5 mag}\right)^2}
\end{equation}
where $\Delta\theta$ is the sky angular separation in arcseconds between the (ra,dec) coordinates listed in SPM4 and IRSF, 
and $\Delta$J is the difference between the IRSF J magnitude and the 2MASS-transformed-to-IRSF one\footnote{Within the SPM4 
LMC catalog, a few hundreds of stars did not have 2MASS JHK but instead
had a value of 0 magnitudes in each band. For these SPM4 stars, we took $\Delta$J$=0$ by default, 
but chose a more restricted $\Delta\theta<2"$. We then checked for repeated matches between this subset of SPM4 stars
and the subset with 2MASS JHKs measured, and found eleven cases where two different
SPM4 stars were matched to the same IRSF source, in each case one match had 2MASS JHKs and the other had 0 values,
and in these few cases, we always chose the one with measured 2MASS JHKs.}. 
The {\it metric} defined by the above equation finds the IRSF star that is closest in {\it both} position and 
magnitude to the SPM4 one. This procedure should minimise false matches.

Having constructed this SPM4$\times$IRSF catalog, one per Magellanic Cloud, 
we proceeded to cross-match the corresponding Be star candidates with them,
this time using only the sky coordinates to find their match:
for each Be star candidate, we found the closest SPM4$\times$IRSF source within 3".

During the matching process in the LMC, 
we found twenty six cases where we had two Be star candidates matched within 3" with one SPM4xIRSF star.
Among these multiple matches, the largest difference in colors
B-V and V-I was about 0.2 magnitudes and generally smaller than 0.05 mags.
Not having other information to decide, we chose the closest in angular separation 
as the most probable true match. We finally obtained 1188 Be star candidates with
SPM4 proper motions and IRSF JHKs magnitudes.

During the matching process in the SMC,
we found nine cases where we had two Be star candidates matched within 3" with one SPM4xIRSF star,
and in one case we had one Be star candidate matched within 3" with two SPM4xIRSF stars.
Among these multiple matches, the largest difference in colors
(B-V and V-I for the first nine cases, and magnitude J for the last one), 
was as observed in the LMC, and not having other information to decide, 
we again chose the closest in angular separation 
as the most probable true match. We finally obtained 619 Be star candidates with
SPM4 proper motions and IRSF JHKs magnitudes.

\begin{figure}
    \includegraphics[width=\columnwidth]{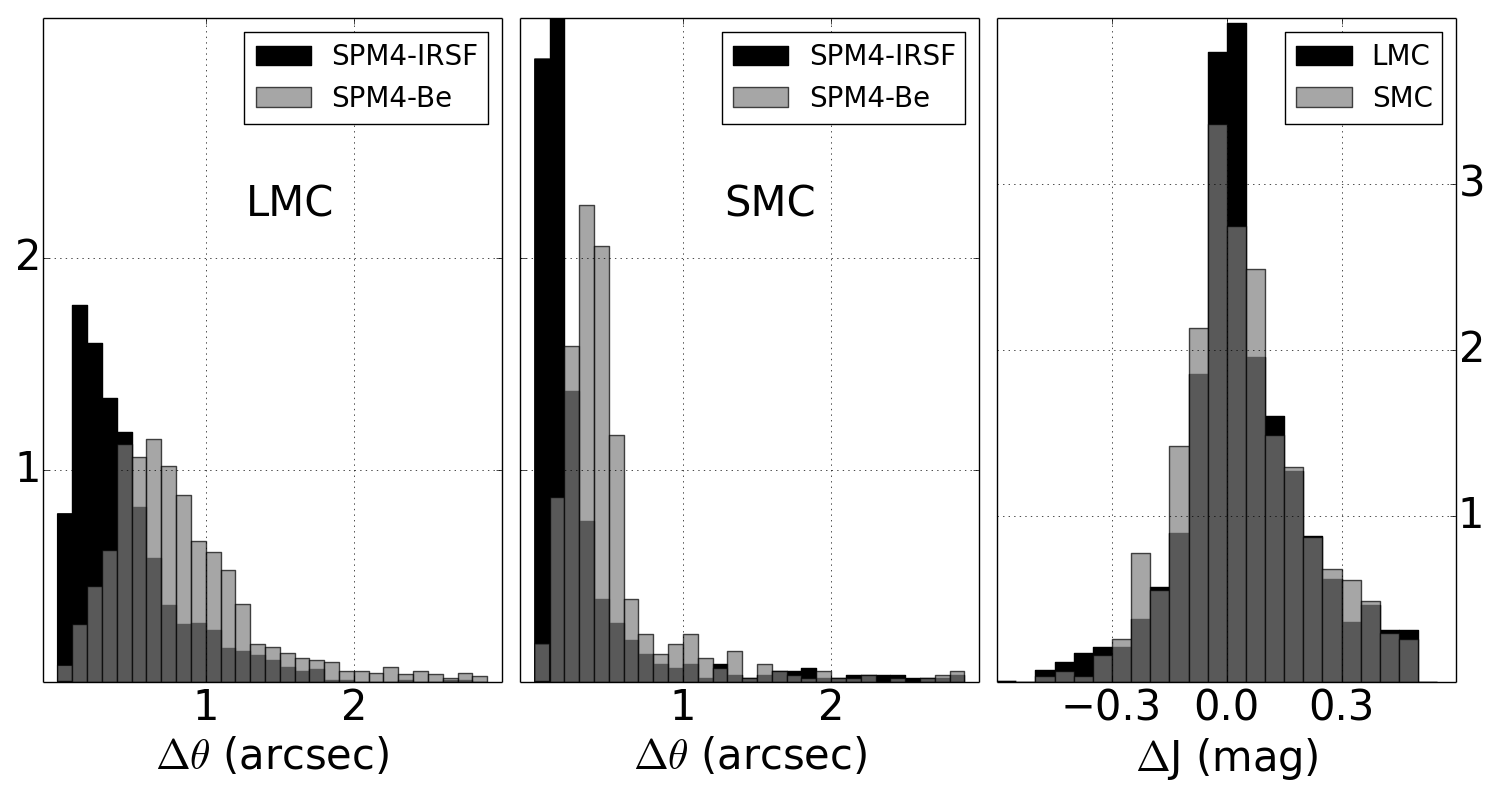}
    \caption{Histograms of $\Delta\theta$ and $\Delta$J for Be star candidates in the LMC and SMC.
     The angular separation between the Be star candidate OGLE-II sky position and the one obtained from SPM4 is also plotted in grey
     in the first two panels. Notice how SPM4 and IRSF have a better agreement than SPM4 and OGLE-II.}
    \label{fig_histo_dtheta_dj}
\end{figure}

Figure \ref{fig_histo_dtheta_dj} shows the histograms of $\Delta\theta$ and $\Delta$J for each of these final matches catalogs,
labeled as SPM4-IRSF. In the first two panels, we also include a histrogram of the sky angular separation in arcseconds between the SPM4 sky coordinate and
the one listed by \citet{sabogal05} and \citet{mennickent02} (taken from OGLE-II), for the LMC and SMC, respectively, labeled as SPM4-Be.
Table \ref{tab:sample_lmc_final_cat} (see Appendix) shows a few lines of the final data table for the LMC sample; the SMC table has the same structure.
Sky equatorial coordinates and proper motions with their corresponding errors come from SPM4, 
BVI photometry comes originally from OGLE-II and JHKs photometry comes from IRSF, the OGLE-II ID as well as the Type
of each Be star candidate comes from \citet{sabogal05} and \citet{mennickent02}, for the LMC and SMC, respectively.
Additional information regarding the quality of the proper motion (columns m, n and i from SPM4), 
as well as the SPM4 ID are included, to identify a few stars in SPM4 whose proper motions are of lower quality
due to the fact that the 2nd-epoch position was not measured neither in SPM plates (n=0) nor CCDs  (i=0) but
was taken from an input catalog. Such few stars are labeled as {\it false} SPM4 stars, as their proper motions
are not measured exclusively with SPM4 data, although their first epoch material is from SPM4 (m indicates
the number of 1st-epoch SPM4 plates used for each star, and for all SPM4 stars, always m$>0$.).

\section{Sample selection from color-color diagrams}
\begin{figure*}
	\includegraphics[width=\textwidth]{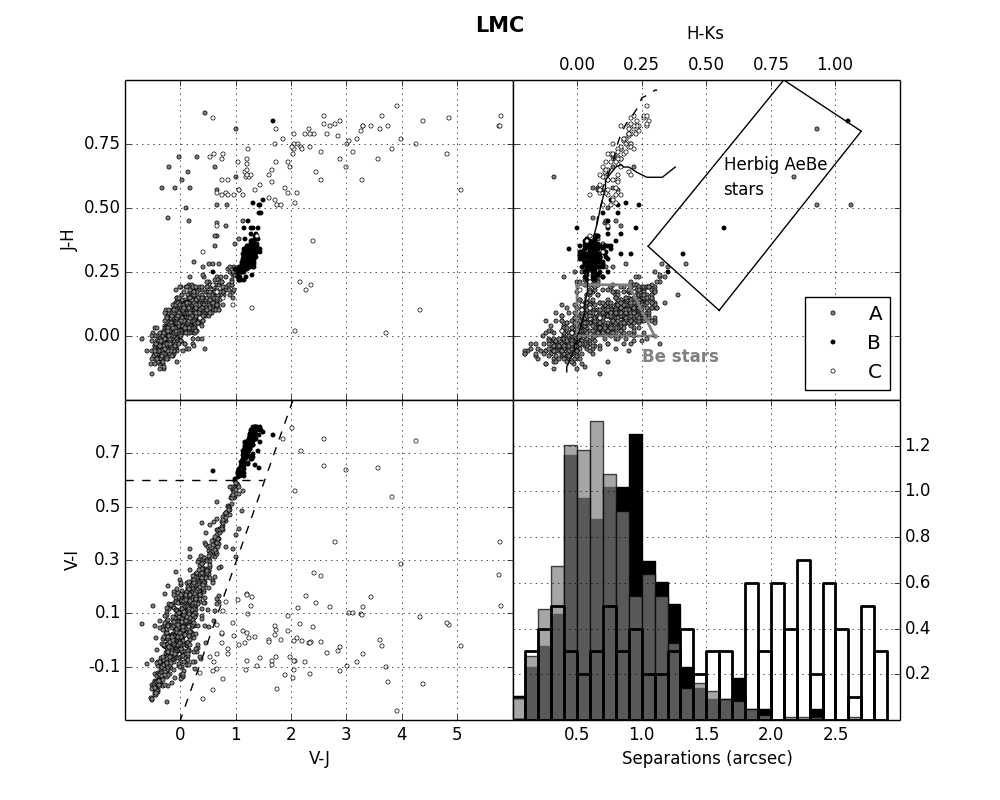}
    \caption{Color-color diagrams for the LMC Be star candidates with proper motions in SPM4 and
    histogram of angular separation between their OGLE-II and SPM4 sky positions, for samples A (gray), B (black), and C (white) within. 
    The dashed straight lines in the lower left panel
    mark the photometry cuts that defined the samples. The solid and dashed curves in the upper right panel are the
    tracks of main sequence and giants stars, respectively, from \citet{allen}. The grey polygon marks the location of Be stars
    from \citet{doug} and the black polygon marks the location of Herbig Ae/Be stars from \citet{jesush}.}
    \label{fig_cc_plot_lmc}
\end{figure*}

\begin{figure*}
	\includegraphics[width=\textwidth]{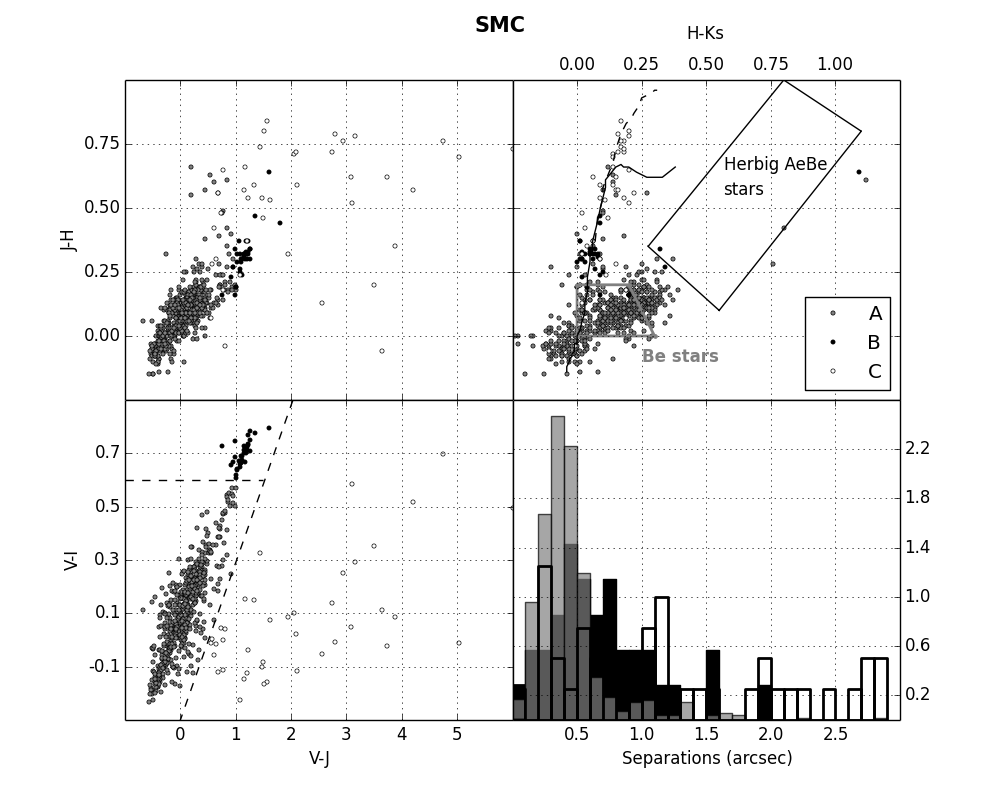}
    \caption{Same than Figure \ref{fig_cc_plot_lmc}, but for the SMC.}
    \label{fig_cc_plot_smc}
\end{figure*}

After close examination of various color-color and color-magnitude diagrams that 
could be done with the collected BVIJHKs photometry, we were able to identify three mutually excluding samples,
labeled A, B and C within each Magellanic Cloud, that had a distinctive clustering in all colours. 
Figures \ref{fig_cc_plot_lmc} (LMC) and \ref{fig_cc_plot_smc} (SMC)
show the results obtained in this regard. 
For both Clouds, these samples were defined with the same cuts in photometry,
as described in Table \ref{tab:sample_def}. The number of {\it false} SPM4 stars within each
sample is shown in parenthesis. 

\begin{table}
	\centering
	\caption{Photometry cuts that define samples A, B and C within each Magellanic Cloud. The number of so called
	 {\it false} SPM4 stars within each sample is written in parenthesis.}
	\label{tab:sample_def}
	\begin{tabular}{clcc} 
		\hline
		Sample & Definition & LMC & SMC\\
		\hline
		A & V-I$\leq$ 0.6 and V-I>0.59(V-J)-0.3 & 864 (15) & 544 (4) \\
		B & V-I$>$0.6  and V-I>0.59(V-J)-0.3 & 216 (3) & 35 (1) \\
		C & V-I$\leq$ 0.59(V-J)-0.3 & 108 (31) & 40 (0) \\
		\hline
		Total & & 1188 (49) & 619 (5) \\
		\hline
		
	\end{tabular}
\end{table}

Such selection was motivated by the results obtained by \citet{paul12}, where a sample of enigmatic
candidates were obtained in the LMC Be star candidates, although not in the SMC,
described as Type-4 stars with different optical and NIR properties.
These stars ``do not have NIR
excess, show large reddening, but are not located in regions with
high reddening. The reddening corrected magnitudes make them the
brightest and most massive stars in the sample. Detailed spectroscopic
studies are needed to understand these enigmatic candidates'' \citep{paul12}.
Within our study, such subgroup correspond to sample B within each cloud,
and as expected, in the SMC such sample is much less populated.
Our figures \ref{fig_cc_plot_lmc} and \ref{fig_cc_plot_smc} 
replicate figures 3 and 4 from \citet{paul12}, although no extinction correction was
included, as it was deemed to small to have any significant effect in our investigation .

The histograms of $\Delta\theta$ shown in the lower right panel of figures \ref{fig_cc_plot_lmc}
and \ref{fig_cc_plot_smc} reveal that samples C within each Magellanic Cloud 
are in fact mismatches between the Be star candidates
and the SPM4$\times$IRSF catalog. Stars in samples C have quite faint V magnitudes 
and J-H and H-Ks colors typical of GKM giants,
as seen in the upper right panels of figures \ref{fig_cc_plot_lmc}
and \ref{fig_cc_plot_smc}. Therefore, samples C within each Magellanic Cloud
are discarded at this point from any further consideration, and only samples A and B 
are analysed in the following sections.

\section{Proper Motion Results}
\begin{figure}
	\includegraphics[width=\columnwidth]{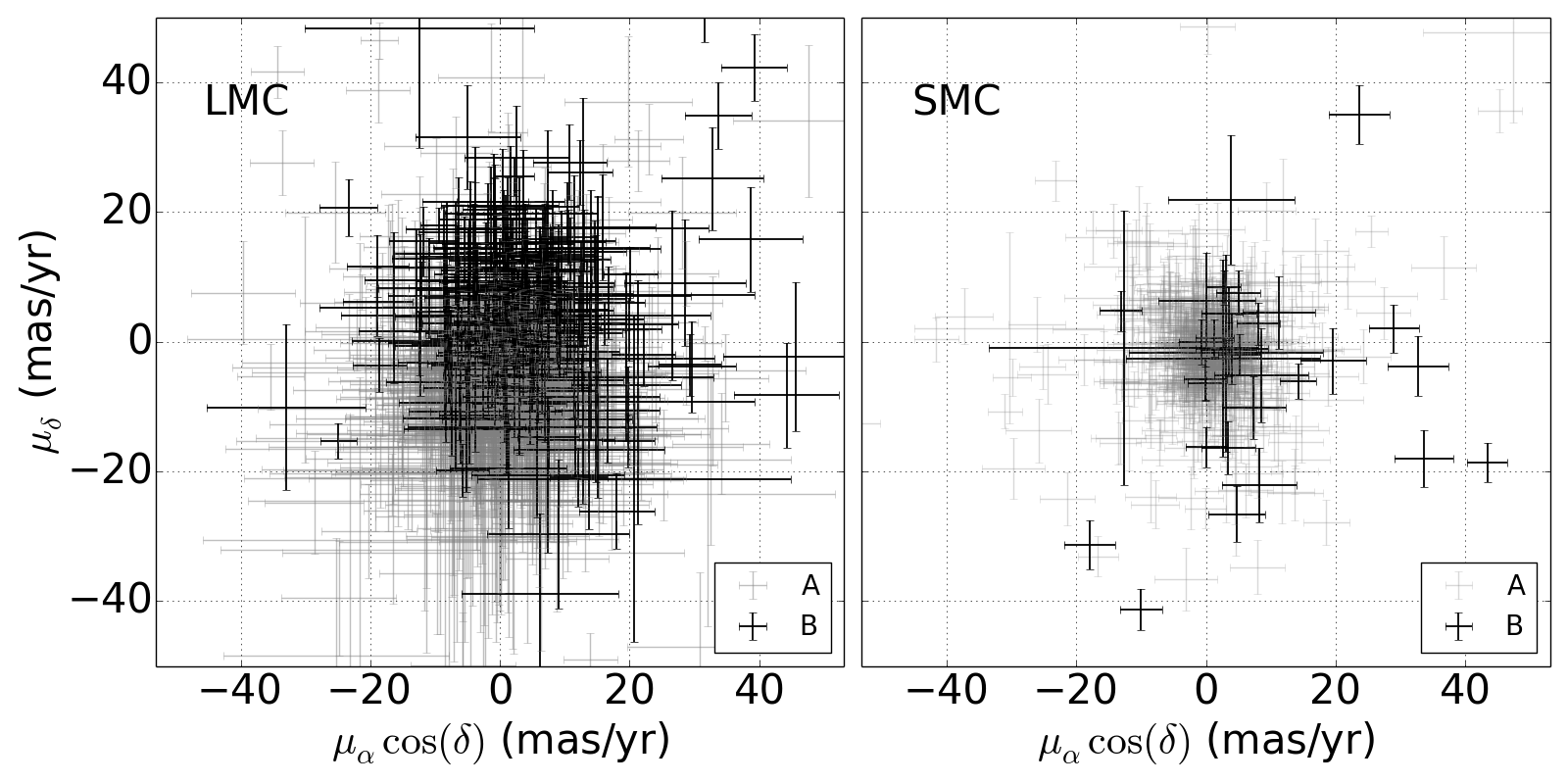}
    \caption{Vector-point diagrams for samples A and B within the LMC (left) and SMC (right).
    In each case, stars in sample B have proper motions too large to belong to the Magellanic
    Clouds and are therefore more probable members of the Milky Way.}
    \label{fig_vpd}
\end{figure}

Figure \ref{fig_vpd} shows the vector points diagrams for samples A and B within each of the Clouds. 
In both cases, although especially for the LMC, 
it is evident that sample B is dominated by stars with large enough
proper motions that place many of its stars within the Milky Way. 
This settles the nature of the originally thought as a new subclass of Be star candidates in the LMC 
suggested by \citet{paul12},
as simply stars located within the Milky Way foreground, having noticeable larger values of $\mu_\delta$
than those expected for stars in the LMC.
In the SMC, we also found that many stars in sample B show systematic larger values, 
this time easily visible in $\mu_{\alpha}\cos(\delta)$,
when compared to the bluer sample A. 
SMC's sample B is of a much smaller size than LMC's sample B.

\begin{figure}
	\includegraphics[width=\columnwidth]{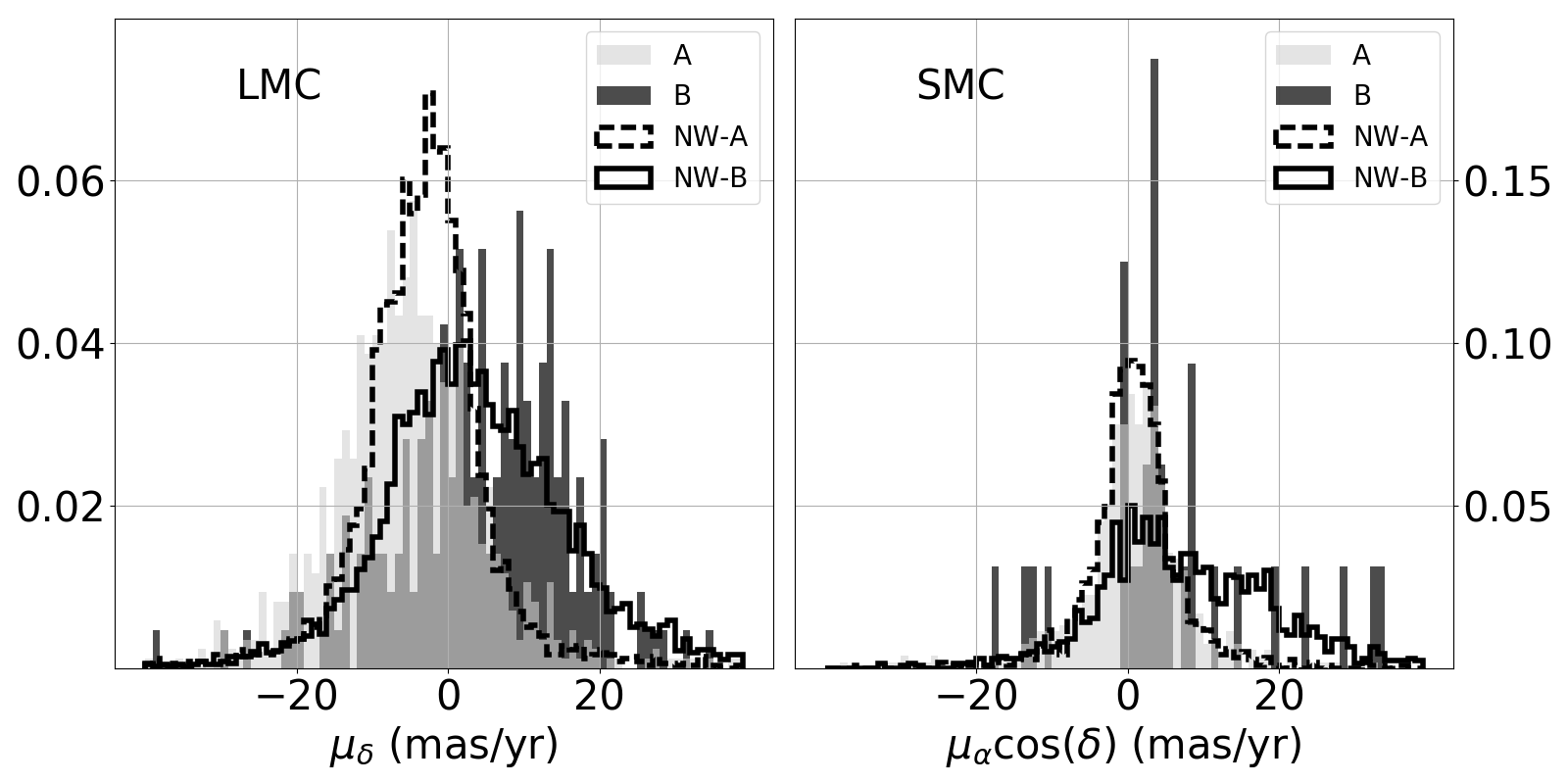}
    \caption{Histograms of $\mu_\delta$ and $\mu_{\alpha}\cos(\delta)$ for the LMC and SMC,
    respectively, for their corresponding samples A and B,  
    and for two samples chosen within each Cloud, and named NW-A and NW-B.}
    \label{fig_histo_pm}
\end{figure}

This result is further confirmed when comparing the histograms of proper motions
for all these samples with ones obtained from the SPM4$\times$IRSF catalog,
using the photometric selection proposed by \citet{niko00} based on JHKs magnitudes.
Figure \ref{fig_histo_pm} shows the histograms of $\mu_\delta$ and $\mu_{\alpha}\cos(\delta)$
for the LMC and SMC, respectively, for their corresponding samples A and B, 
and for two samples chosen within each Magellanic Cloud, and named NW-A and NW-B respectively, defined as described
in Table \ref{tab:NW-sample_def}. Samples NW-A are dominated by stars in the Magellanic Clouds 
of spectral type and luminosity class O3-O5 V and B-A I-II, while samples NW-B are dominated
by Milky Way disk stars of spectral type and luminosity class F-K V.
It is important to notice that to select these NW-samples, we used IRSF JHKs magnitudes.
It is visible that samples A and NW-A follow similar kinematics, and so do samples B and NW-B,
within each Magellanic Cloud. This similarity refers mainly to central tendency values,
that is that the mean values of samples A and NW-A are close to each other, and so are in samples B and NW-B,
although regarding dispersion, each pair of samples also bear resemblance.
Table \ref{tab:NW-stats} contains the number of data points, 
mean and unbiased standard deviation of  $\mu_\delta$ for LMC,
and  $\mu_{\alpha}\cos(\delta)$ for the SMC, for their corresponding samples. 

In any case, the main point of this investigation is to prove that samples B do not follow the kinematics
expected for the Ma\-ge\-lla\-nic Clouds, that is that samples B proper motion distribution is noticeably
different from their corresponding samples NW-A. 
Two-samples Welch's T-tests were performed on each possible pair of samples
to test the null hypothesis that the means of the compared samples are the same. 
The obtained p-values are in Table \ref{tab:NW-welch} and indicate that the null hypothesis
should be clearly rejected when comparing samples B vs. NW-A, and  A vs. NW-B ones.
These tests also show that the null hypothesis should not be indisputably rejected in the other cases,
that is A vs. NW-A and B vs. NW-B, except on the LMC A vs. NW-A case, where it is visible that some difference 
exists in the means. Still in this case, we notice that their distributions overlap well enough to
substantiate the idea that A and NW-A share a similar kinematics, i.e. that many of the Be star
candidates in the LMC sample A are indeed located in the LMC. It should be taken into
account that SPM4 proper motion errors do affect the performance of these tests, and contamination
from other populations or the presence of systematic effects in the SPM4 proper motions should
be considered (See section \ref{sec_discussion} for an analysis of this last point).


\begin{table}
	\centering
	\caption{Photometry cuts that define samples NW-A and NW-B within each Magellanic Cloud,
	following \citet{niko00} recipes for the LMC. For the SMC, the cut in Ks magnitude was extended
	  0.5 magnitudes fainter to account for its larger distance modulus, the color selection was kept the same.}
	\label{tab:NW-sample_def}
	\begin{tabular}{crcl} 
		\hline
		Sample & \multicolumn{3}{c}{Photometric cut} \\
		\hline
		LMC NW-A & Ks<14.75 & and & J-Ks<0.2 \\
		LMC NW-B & Ks<13.5 & and & 0.2<J-Ks<0.5  \\
		\hline
		SMC NW-A & Ks<15.25 & and & J-Ks<0.2  \\
		SMC NW-B & Ks<14 & and & 0.2<J-Ks<0.5  \\
		\hline
	\end{tabular}
\end{table}



\begin{table}
	\centering
	\caption{Basic statistics for samples A, NW-A, B and NW-B in the LMC and SMC. SD stands for unbiased standard deviation.}
	\label{tab:NW-stats}
	\begin{tabular}{crrr} 
		\hline
		Sample & Size & Mean & SD \\
		\hline
		\multicolumn{2}{c}{LMC} & \multicolumn{2}{c}{$\mu_\delta$}   \\
		\hline
		A & 864 & -5.64 & 11.57 \\
		NW-A & 3621 & -3.12 & 11.41 \\
		B & 216 & 4.99 & 12.63 \\
		NW-B & 4838 & 3.86 & 13.24  \\
		\hline
		\multicolumn{2}{c}{SMC} & \multicolumn{2}{c}{$\mu_{\alpha}\cos(\delta)$}  \\
		\hline
		A & 544 & 1.44 & 13.49 \\
		NW-A & 1202 & 1.18 & 11.04 \\
		B & 35 & 10.06 & 18.45  \\
		NW-B & 1397 & 8.29 & 14.07  \\
		\hline
	\end{tabular}
\end{table}

%
%

\begin{table}
	\centering
	\caption{Welch's T-test results (p-values) for all possible comparisons of samples A, NW-A, B and NW-B in the LMC and SMC.}
	\label{tab:NW-welch}
	\begin{tabular}{ccccc} 
		\hline
		& \multicolumn{2}{c}{LMC} & \multicolumn{2}{c}{SMC}   \\
		& A & B & A & B \\
		\hline
		NW-A & $1.01\times 10^{-8}$ &  $1.86\times 10^{-17}$ & 0.91 & $5.38\times 10^{-3}$ \\
		NW-B & $1.85\times 10^{-89}$ & 0.20 & $5.75\times 10^{-11}$ & 0.18 \\
		\hline
	\end{tabular}
\end{table}

\subsection{Cross-examination with some measured radial velocities}
In \citet{paul12}, radial velocities and spectral types were determined for 102 Be star candidates
drawn from \citet{mennickent02} and \citet{sabogal05}, 39 and 63 stars in the LMC and SMC,
respectively. In the SMC list, the star with OGLE-II ID 005224.40-724038.6
appeared twice, with the same sky coordinates and photometry but with Types 1 and 2, and different radial velocities
though still consistent with each other within errors ($V_r=149\pm 30$ km/s and $V_r=115\pm24$ km/s, respectively).
In fact this star was identified in our work and the Type-1 entry was taken for the match with SPM4,
so in this case, for consistency we do the same again.
From the LMC sample, 8 stars (out of 39) were identified in our work with proper motions in SPM4 
and their data are displayed in Table \ref{tab:lmc_rv}.
From the SMC sample, 33 stars (out of 63) were identified in our work with proper motions in SPM4 
and the data for a few stars are shown in Table \ref{tab:smc_rv}.

\begin{figure}
	\includegraphics[width=\columnwidth]{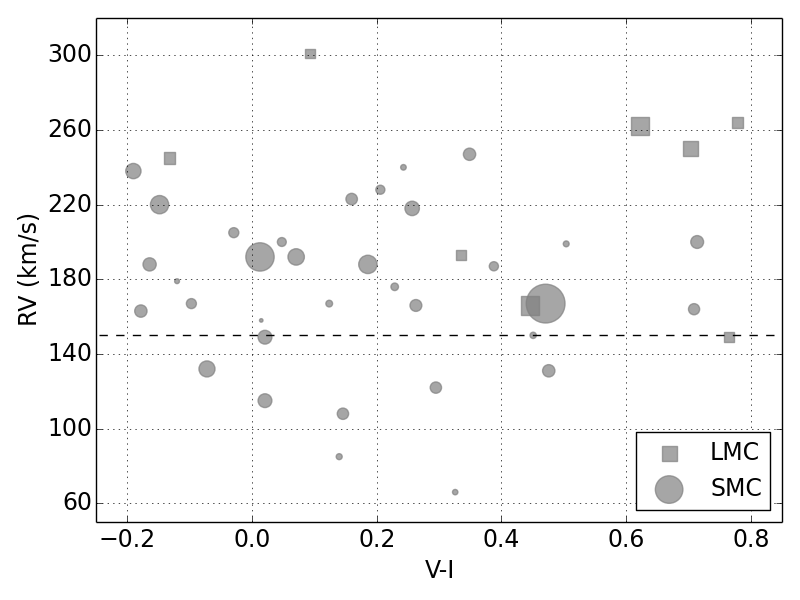}
    \caption{Radial velocity for Be stars with SPM4 proper motions versus their OGLE-II $V-I$ color, 
    with a symbol sized by the total proper motion $\mu=\sqrt{(\mu_{\alpha}\cos(\delta))^2+(\mu_\delta)^2}$.}
    \label{fig_rv_vi}
\end{figure}

Figure \ref{fig_rv_vi} shows the radial velocity RV for these stars versus the V-I color\footnote{Be star candidate 
with OGLE-II ID 004526.51-733014.2 in the SMC has V-I=1.092, out of the plot range and therefore is not shown in this figure. 
The same applies for Figure \ref{fig_pm_vi}.}, 
with a symbol sized by the total proper motion $\mu=\sqrt{(\mu_{\alpha}\cos(\delta))^2+(\mu_\delta)^2}$.
Despite the fact that proper motion errors at the distance of the Magellanic Clouds represent a much larger
error in km s$^{-1}$ as compared to the RV errors, it can be noticed how many of the SMC bluer stars 
have RV>150 km s$^{-1}$ and also the smaller proper motions, therefore all kinematical information
is consistent with their membership to the SMC. For the LMC stars, with so few stars this plot
is not as useful. In any case, this plot confirms that the bluer a variability-selected Be star candidate is
towards the Magellanic Clouds, more probable it is to belong to these galaxies.

\section{Discussion}\label{sec_discussion}

When finding trends of proper motions among colour-selected samples, 
it is important to check that such behaviour is not the result of systematic
errors within the catalog used, and therefore of non-cosmic origin, so to say.

In the Magellanic Clouds area, the SPM4 catalog has measured proper motions,
using observational materials of different kinds for their 2nd-epoch positions:
photographic plates or CCDs (in a very few cases, these positions were directly taken from an input catalog, 
which can be Hipparcos, Tycho-2, UCAC2 or 2MASS, then $n=i=0$). Proper motions
computed from 2nd-epoch CCD have errors about 2-3 times smaller
than those from 2nd-epoch plates, and such errors are correlated with 
position in the sky, depending on the location of the plates. It can also happens
that from one area in the sky with plates to another one with CCDs,
an offset in the reference system takes place, artificially introducing 
trends in proper motions. Many astrometric catalogs suffer from these
problems, but a thorough examination of the results generally reveals
if such issues are present.

A quick examination of the proper motions in the Magellanic Clouds area
within the SPM4 catalog, reveals some trends with $n$ that were inspected 
carefully to check if those could be behind the results described in the previous section.
No trends in proper motion with $m$ or $i$ were observed,
which in a catalogue like SPM4, are not unexpected.

\begin{figure}
	\includegraphics[width=\columnwidth]{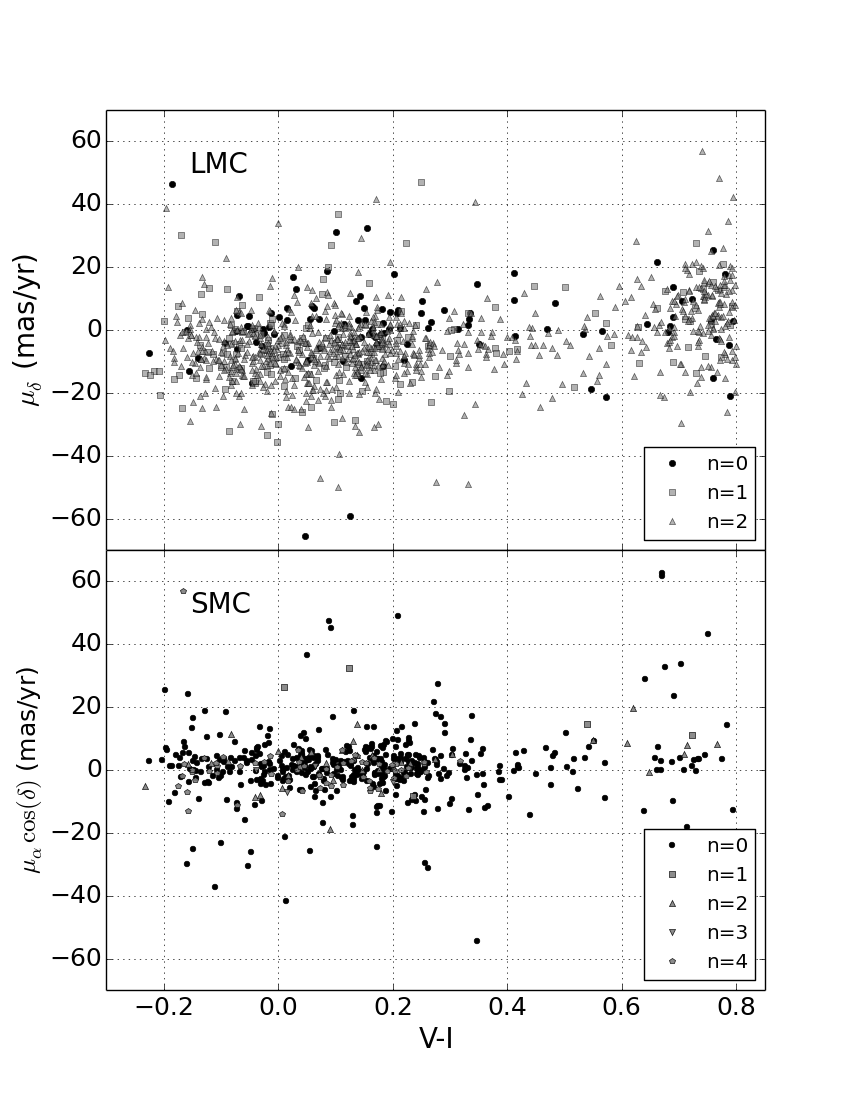}
    \caption{Proper motion vs. V-I, for different values of $n$.}
    \label{fig_pm_vi}
\end{figure}

Within the LMC SPM4$\times$IRSF catalog, it is visible that stars with $n=0$
(no 2nd-epoch plates, therefore CCD positions if $i>0$) have an overall shift in their 
$\mu_\delta$ values towards positive values, as compared to the stars with $n>0$,
whose global distribution looks more centered towards zero.
Within the LMC Be sample, about 10\%, 20\% and 70\% of the stars have $n=0$, 
$n=1$ and $n=2$, respectively. The proper motion trend in $\mu_\delta$ is
the same for $n=1,2$ and as described in the previous section,
and for $n=0$ is a bit offset but still sample B shows positive and larger
values in $\mu_\delta$ as compared to sample A,
as shown in Fig. \ref{fig_pm_vi}, upper panel.
Within samples A and B,  $n=1, 2$ stars represent 73\% and 83\% of the stars, respectively. 
The majority of the $n=0$ stars are located towards the west end of the LMC Be stars in the sky.
Within the SMC SPM4$\times$IRSF catalog, no visible trend in the proper motions
with $n$ was visible. Within the SMC Be sample, about 80\% of the stars have $n=0$,
and the same proportion is seen within sample A and sample B. 
Figure \ref{fig_pm_vi}, lower planel,  shows $\mu_\alpha\cos(\delta)$ vs. $V-I$
color-coded by $n$, which in this case ranges from 0 to 4.
We conclude that the observed trends in proper motions are a real feature of the
samples chosen and not a bias caused by systematics in the SPM4 catalogue.

\section{Conclusions}
\begin{itemize}
\item A proper motion investigation for a sample of Be star candidates towards the Magellanic Clouds,
using data from the SPM4 catalog has been done, which has allowed us to separate a Galactic foreground population
from the Magellanic background one. 
\item The Galactic foreground population has redder colors in B-V and V-I and large proper motions visibly
different from those expected for the Magellanic Clouds.
\item In the LMC, this population is readily visible as a cluster in the J-H vs. J-Ks color-color diagram and had been
described in the literature as an intriguing subgroup of stars within the LMC, that needed further spectroscopic analysis to clarify their nature. 
\item Our result proves that these stars are simply not in the Magellanic Clouds and confirms
that the photometric variability method used by \citet{mennickent02} and \citet{sabogal05}  to select Be star candidates, 
suffers from significant contamination towards redder B-V and V-I values.
\end{itemize}

%








\appendix

\section{Tables}
A sample of Table \ref{tab:sample_lmc_final_cat}, Table \ref{tab:lmc_rv} and a sample of Table \ref{tab:smc_rv} are shown in the following page.

\bsp	

\begin{landscape}
\begin{table}
	\centering
	\caption{Sample of the final table for the Be stars candidates in the LMC, that includes OGLE-II BVI and IRSF JHKs photometry and
	 SPM4 proper motions for each star. The SMC table has the same structure.  Sky positions are given in degrees, proper motions and their
	 errors are in mas yr$^{-1}$.}
	\label{tab:sample_lmc_final_cat}
	\begin{tabular}{cccccccccccccccccc} 
	OGLE-II ID &    V &   B-V  &  V-I   & Type  & SPM4 ID   & $\alpha$ & $\delta$  & $\mu_\alpha\cos(\delta)$ & $\mu_\delta$ &  
	$\epsilon_{\mu_\alpha\cos(\delta)}$ & $\epsilon_{\mu_\delta}$ & m & n & i &  J &   H &   Ks \\  \hline
  05263636-6932003 & 15.12  & -0.098 & 0.038  & 4    & 590032780 & 81.6505226 & -69.5335014 & -15.12 & -11.5  & 8.04  & 8.06  & 2 & 2 & 0 & 15.034 & 14.928 & 14.904 \\
  05390419-7005011 & 16.499 & -0.172 & 0.049  & 4    & 590488238 & 84.7670294 & -70.0836831 & 9.1    & -9.54  & 11.96 & 11.7  & 2 & 2 & 0 & 16.003 & 15.76  & 15.707 \\
  04595836-6925494 & 15.949 & 0.007  & 0.185  & 1    & 580392707 & 74.992886  & -69.430346  & 6.07   & -1.08  & 3.53  & 3.54  & 3 & 0 & 2 & 15.737 & 15.418 & 15.069 \\
  05013017-6908270 & 16.058 & -0.028 & 0.126  & 1    & 580452163 & 75.375393  & -69.1403308 & 8.53   & -59.09 & 6.11  & 6.0   & 4 & 0 & 1 & 15.632 & 15.331 & 13.904 \\
  05064836-6837230 & 15.573 & -0.015 & 0.145  & 1    & 580542816 & 76.7010587 & -68.6228068 & -6.82  & -11.6  & 6.02  & 5.94  & 4 & 2 & 0 & 15.293 & 14.969 & 14.946 \\
  05065094-7000527 & 16.11  & -0.085 & -0.067 & 1    & 580290171 & 76.7123547 & -70.0147158 & 23.44  & -11.82 & 8.19  & 8.13  & 4 & 2 & 0 & 16.375 & 15.428 & 15.277 \\
  05065274-6834374 & 15.698 & -0.016 & 0.158  & 1    & 580033854 & 76.7191622 & -68.5770339 & -4.5   & -11.75 & 5.7   & 5.61  & 4 & 1 & 0 & 15.37  & 15.267 & 14.93  \\
  05070447-6847599 & 18.654 & 0.072  & 0.088  & 1    & 580033896 & 76.7665291 & -68.8001924 & -1.49  & -14.38 & 4.71  & 4.73  & 4 & 2 & 0 & 14.316 & 14.225 & 14.203 \\
  05013061-6838567 & 16.346 & 0.012  & -0.028 & 1    & 580540617 & 75.3773197 & -68.6489925 & -8.79  & -4.94  & 8.9   & 8.61  & 4 & 0 & 1 & 16.396 & 15.517 & 15.597 \\
  05071953-6845149 & 14.733 & -0.19  & -0.168 & 1    & 580033947 & 76.8312574 & -68.7542808 & 7.06   & -24.67 & 7.12  & 7.06  & 3 & 1 & 0 & 15.179 & 15.268 & 16.444 \\ \hline
	\end{tabular}
\end{table}

\begin{table}
	\centering
	\caption{Be stars candidates in the LMC that have full kinematic information, e.g. radial velocity and proper motions.
	Radial velocities and their errors, RV and e\textsubscript{RV}, are given in km s$^{-1}$, proper motions and their
	 errors are in mas yr$^{-1}$.}
	\label{tab:lmc_rv}
	\begin{tabular}{cccccccccc} 
OGLE-II ID &  Spectral Type &  RV & e\textsubscript{RV} & V-I &  SPM4 ID & $\mu_\alpha\cos(\delta)$ & $\mu_\delta$ & 
$\epsilon_{\mu_\alpha\cos(\delta)}$ & $\epsilon_{\mu_\delta}$ \\ \hline
  050052.36-685803.7 & F9       & 149 & 2   &  0.765  & 580033194 & 4.62   & -2.98 &  2.19 & 2.24 \\
  050303.70-690615.0 & A9-F0  & 193 & 33  &  0.335  & 580027763 & -1.84 &  5.28  &  2.39 & 2.44 \\
  051647.54-694415.2 & F3       & 264 & 33   & 0.779  & 580023041 & 0.52   & -6.85 &  4.74 & 4.76 \\
  051744.42-692033.3 & B8-A1  & 301 & 13  &  0.093  & 580029761 & 3.92   & -3.1   & 3.38 & 3.44 \\
  051947.82-693912.3 & F6       & 166 & 43  &  0.446  & 580023530 & -11.34 & 13.93 &  4.98 & 5.04 \\
  052032.26-694224.2 & F6       & 262 & 20  &  0.622  & 580346495 & 7.47   & -14.68 & 4.8  & 4.81 \\
  052258.47-692621.0 & F6       & 250 & 99   & 0.703  & 580030795 & 4.67   & -10.82 & 4.87 & 4.89 \\
  052402.01-694920.5 & B2-B3  & 245 & 22   & -0.132 & 580024625 & 5.89   & 3.35   & 4.4  & 4.47 \\ 
  \hline
	\end{tabular}
\end{table}

\begin{table}
	\centering
	\caption{Be stars candidates in the SMC that have full kinematic information, e.g. radial velocity and proper motions.
	Radial velocities and their errors, RV and e\textsubscript{RV}, are given in km s$^{-1}$, proper motions and their
	 errors are in mas yr$^{-1}$.}
	\label{tab:smc_rv}
	\begin{tabular}{cccccccccc} 
OGLE-II ID &  Spectral Type &  RV & e\textsubscript{RV} & V-I &  SPM4 ID & $\mu_\alpha\cos(\delta)$ & $\mu_\delta$ & 
$\epsilon_{\mu_\alpha\cos(\delta)}$ & $\epsilon_{\mu_\delta}$ \\ \hline
005224.40-724038.6 & B0-B3 & 149 & 30& 0.021 & 290026279 & -5.77 & 8.01 & 2.85 & 2.94 \\
004554.14-731404.3 & A7-F0 & 164 & 37& 0.709 & 280020464 & -0.06 & -6.46 & 2.63 & 2.71 \\
004750.14-731316.4 & B8 & 167 & 69& 0.471 & 290213545 & -77.67 & -7.86 & 4.08 & 4.21 \\
004921.41-725844.9 & A7-F0 & 199 & 30& 0.504 & 290025783 & 0.81 & -1.59 & 4.14 & 4.26 \\
005025.64-725807.1 & A7-F0 & 150 & 29& 0.451 & 290025912 & -1.4 & -1.5 & 3.95 & 4.05 \\
005043.44-732705.3 &  & 192 & 58& 0.013 & 290190227 & -41.61 & 0.36 & 3.39 & 3.45 \\
005359.22-723508.9 &  & 66 & 23& 0.326 & 290026564 & -0.12 & -1.5 & 3.14 & 3.23 \\
005745.25-723532.1 & B2-B5 & 220 & 23& -0.148 & 290027028 & 16.43 & 5.42 & 2.36 & 2.43 \\
010000.78-725522.9 &  & 240 & 24& 0.243 & 290027215 & 0.32 & -1.58 & 3.82 & 3.93 \\
010043.94-722604.8 & B0-B3 & 188 & 14& -0.164 & 290032264 & 8.91 & -1.62 & 2.07 & 2.13 \\ \hline
\end{tabular}
\end{table}

\end{landscape}


\label{lastpage}
\end{document}